\documentclass[prd, twocolumn, nofootinbib, floatfix]{revtex4-1}

\usepackage{amsmath}
\usepackage{graphicx}
\usepackage{dcolumn}
\usepackage{bm}
\usepackage{epsfig}
\usepackage{amssymb,latexsym,mathrsfs}
\usepackage{graphicx}
\usepackage{color}
\usepackage{hyperref}

\hypersetup{
    colorlinks=true,
    linkcolor=red,
    citecolor=blue,
} 

\newcommand{\be}{\begin{equation}}
\newcommand{\ee}{\end{equation}}
\newcommand{\bs}{\begin{split}} 
\newcommand{\bea}{\begin{eqnarray}}
\newcommand{\eea}{\end{eqnarray}}

\newcommand{\om}{\Omega_m}

\newcommand{\hiso}{H_{\rm iso}}
\newcommand{\aiso}{a_{\rm iso}}
\newcommand{\ode}{\Omega_{de}}
\newcommand{\odei}{\Omega_{de,iso}}

\begin{document}

\title{Probing Dark Energy Anisotropy} 
\author{Stephen A.\ Appleby$^1$ \& Eric V.\ Linder$^{1,2}$} 
\affiliation{$^1$ Institute for the Early Universe WCU, Ewha Womans 
University, Seoul, Korea} 
\affiliation{$^2$ Berkeley Lab \& University of California, Berkeley, 
CA 94720, USA}

\begin{abstract}
Wide area cosmological surveys enable investigation of whether dark energy 
properties are the same in different directions on the sky.  Cosmic 
microwave background observations strongly restrict any dynamical effects 
from anisotropy, in an integrated sense.  For more local constraints we 
compute limits from simulated distance measurements for various 
distributions of survey fields in a Bianchi I anisotropic universe.  
We then consider the effects of fitting for line of sight properties 
where isotropic dynamics is assumed (testing the accuracy through 
simulations) and 
compare sensitivities of observational probes for anisotropies, from 
astrophysical systematics as well as dark energy.  
We also point out some interesting features of anisotropic expansion in 
Bianchi I cosmology. 
\end{abstract}

\date{\today} 

\maketitle

%%%%%%%%%%%%%%%%%%%%%%%%%%%%%%%%%%%%%%%%%%%%%%%%%%%%%%%%%%%%%%%%%%%%%%%%
\section{Introduction} 

The time variation of the cosmic expansion gives key clues to the 
energy components of the universe, with the acceleration pointing to 
an unknown dark energy.  As cosmological surveys cover more of the 
sky in more detail we can also examine spatial variation of the 
expansion and dark energy properties.  Here we investigate anisotropy 
rather than inhomogeneities.  While the cosmic microwave 
background radiation places tight constraints on any anisotropy, 
ensuring a close to isotropic global expansion, smaller scale pressure 
anisotropies 
that do not disrupt the global isotropy remain possible.  In particular 
these can also arise from astrophysical systematics, but we can phrase this 
in terms of variations in the effective dark energy pressure, and explore 
detectable signatures. 

In testing for anisotropy or consistency with isotropy we can ask which 
cosmological probes are 
most sensitive in what redshift ranges to such a hypothetical anisotropy, 
i.e.\ what constraints could be put on angular variations in the local dark 
energy equation of state.  The dark energy equation of state, which can 
also be interpreted in terms of an anisotropic pressure, is of interest 
because of its close connection with fundamental properties of the 
physics behind dark energy.  As we will see, it also gives close 
connections with exact solutions of anisotropic spacetimes such as 
Bianchi models. 

Other work has explored dark energy anisotropy in terms of the small 
scale spatial inhomogeneities in its density \cite{Cooray:2008qn}, large scale 
anisotropies giving an overall ellipticity to the universe 
\cite{Koivisto:2007bp}, and within specific models such as vector dark 
energy  \cite{Koivisto:2008ig,Cooke:2009ws,Pereira:2007yy,Jimenez:2008au,Jimenez:2008nm,Jimenez:2009py,Zuntz:2010jp,thorsrud}, 
elastic dark energy \cite{Battye:2006mb,Battye:2007aa,Battye:1999eq}, 
noncommutativity \cite{07081168}, etc. 
Our approach uses exact solutions, similar to \cite{Appleby:2009za}, 
as well as phenomenological line of sight anisotropy but global isotropy, 
similar to \cite{lsstbook,sullivan}, testing the difference, 
exploring further probes, 
considering sources of astrophysical systematics, and motivating the 
phenomenology with comparisons to exact Bianchi solutions. For early 
and other work on anisotropic spacetimes see 
\cite{Hawking:1968zw,Collins:1972tf,Barrow,barrow97,barrow2010,08060496,10064638}.

In Section~\ref{sec:bianchi} we draw lessons from the exact solutions of 
Bianchi I cosmology to underscore the difficulty of global anisotropy and 
to motivate a possible alternate approach to anisotropic dark 
energy.  We apply the Raychaudhuri beam equation of light propagation 
in Sec.~\ref{sec:ray} and simulate how surveys using, e.g., supernova 
distances in different sky patches could constrain anisotropy.  A line of 
sight anisotropic model reminiscent of the Dyer-Roeder \cite{Dyer:1973zz} 
treatment of inhomogeneities is then investigated in Sec.~\ref{sec:sens} 
to determine the sensitivity of a variety of cosmological probes to 
detecting anisotropic dark energy or astrophysical systematics.  
We conclude in Sec.~\ref{sec:concl}.

%%%%%%%%%%%%%%%%%%%%%%%%%%%%%%%%%%%%%%%%%%%%%%
\section{Exact Solution: Bianchi I Cosmology} \label{sec:bianchi} 

To assess the influence of both the global expansion and the line of 
sight conditions on light propagation we examine an anisotropic exact 
solution of the Einstein field equations.  The Bianchi I cosmology has 
different expansion rates along the three orthogonal spatial directions, 
given by the metric 
\be 
ds^2=-dt^2+a(t)^2 dx_a^2 +b(t)^2 dx_b^2 +c(t)^2 dx_c^2 \ . 
\ee 
The model is homogeneous but anisotropic.  This can arise from a 
homogeneous and isotropic density but anisotropic pressure, for example.  
We can choose the matter and radiation components 
to be isotropic but the dark energy pressure to be different along the 
three axes, with equation of state ratios $w_i=P_i/\rho_{de}$. 

We begin by examining the global dynamics. 
Although the full sky angular average of the dark energy equation of state 
is $\bar w=(w_a+w_b+w_c)/3$, the average expansion rate 
$\bar H=(H_a+H_b+H_c)/3$ does not behave exactly like in an isotropic 
universe with $\bar w$.  To quantify this, define $\hiso$ to be the 
isotropic, Friedmann-Robertson-Walker (FRW) expansion rate for a universe 
with the same present 
matter density (and dark energy density) and with isotropic dark energy 
equation of state $\bar w$.  We can then rewrite the Einstein field 
equations in terms of the ratio $h_i\equiv H_i/\hiso$ and explore the 
deviations from isotropy. 

This gives rise to an autonomous system of equations 
\bea 
h_a'&=&\frac{3}{2}h_a-h_a^2-\frac{1}{2}(h_ah_b+h_ah_c-h_bh_c) 
\label{eq:littlehbegin} \\ 
&&\ -\frac{3}{2}\ode\,(w_b+w_c-w_a)+\frac{3}{2}\bar w h_a\odei \notag\\ 
h_b'&=&\frac{3}{2}h_b-h_b^2-\frac{1}{2}(h_bh_c+h_bh_a-h_ch_a)\\ 
&&\ -\frac{3}{2}\ode\,(w_c+w_a-w_b)+\frac{3}{2}\bar w h_b\odei \notag\\  
h_c'&=&\frac{3}{2}h_c-h_c^2-\frac{1}{2}(h_ch_a+h_ch_b-h_ah_b)\\ 
&&\ -\frac{3}{2}\ode\,(w_a+w_b-w_c)+\frac{3}{2}\bar w h_c\odei \notag\\  
\ode'&=&-\ode\,[(1+w_a)h_a+(1+w_b)h_b+(1+w_c)h_c \notag\\ 
&&\quad\qquad -3-3\bar w\odei] \\ 
\om'&=& -\om\,[h_{a}+h_{b}+h_{c}-3-3\bar{w}\odei]\label{eq:littlehend} \ , 
\eea
where prime denotes $d/d\ln\aiso$.  The isotropic scale factor is used as 
a measure of time; note it is not equal to the monopole anisotropic scale 
factor $\bar a=(abc)^{1/3}$.  The time dependent dimensionless dark energy 
and matter densities $\ode$ and $\om$ are defined as 
$\Omega_i\equiv 8\pi G\rho_i/(3\hiso^2)$, and $\odei$ denotes the dark 
energy density in the isotropic case, with equation of state $\bar w$.  
Numerically we evolve equations (\ref{eq:littlehbegin}-\ref{eq:littlehend}) 
and use the Friedmann-like equation 
\be 
h_{a}h_{b} + h_{a}h_{c} + h_{b}h_{c} = 3(\om+\ode) \label{eq:Fri} 
\ee 
as a consistency check at each timestep. 

\begin{figure}[htbp!]
\includegraphics[width=\columnwidth]{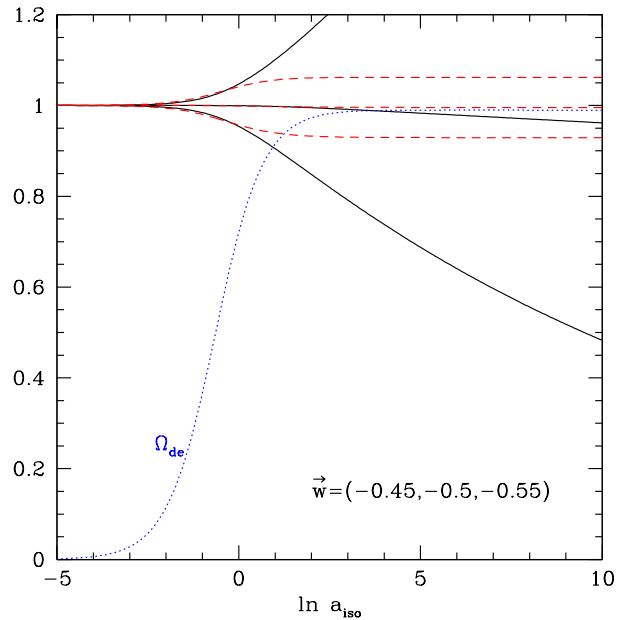}
\caption{Anisotropic expansion of a model with $\vec w=(-0.45,-0.5,-0.55)$ is 
plotted vs $\ln\aiso$ from the early to late universe.  Solid black curves 
give the scale factors, and dashed red curves give the expansion rates, along 
the symmetry axes as ratios to an isotropic, $\bar w=-0.5$ FRW universe. 
Early time and late time fixed points in expansion rate are seen. 
}
\label{fig:abc}
\end{figure}

Numerical solutions to the field equations appear in Fig.~\ref{fig:abc}. 
The early universe appears isotropic, with deviations in the expansion 
rate along symmetry axis $i$ of order $(\bar w-w_i)\ode$.  So when 
$\ode\ll1$ the universe is effectively isotropic.  As the dark 
energy becomes more dynamically important, the anisotropy grows.  However, 
note there is a late time fixed point (for $\bar w>-1$) 
such that the expansion rates go to 
constant offsets from the isotropic behavior.  This is quite interesting: 
the universe does not ``pancake'' in terms of expansion rate 
(although the ellipticity does diverge), but rather it retains 
some memory of the isotropic 
state and remains nearly isotropic in some average sense.

The fixed point solutions can be calculated analytically to various 
orders in the equation of state anisotropy.  Take the dark energy equation 
of state along the three symmetry axes to be 
\be 
(w_a,w_b,w_c)=(\bar w-e-f,\bar w+e,\bar w+f) \ . 
\ee  
Assuming both $e$ and $f$ are small compared to $1+\bar w$, i.e.\ 
$\Delta w\equiv |w_i-\bar w|\ll 1+\bar w$, the 
asymptotic solutions as we approach the limit $a_{\rm iso} \to \infty$ for the 
expansion rates $h_i=H_i/H_{\rm iso}$ normalized to the isotropic rate 
$H_{\rm iso}(\bar w)$ are, to second order, 
\bea 
h_a&=&1-\frac{2(e+f)}{1-\bar w}-\frac{4}{3}\frac{e^2+f^2+ef}{1-\bar w^2} \\ 
h_b&=&1+\frac{2e}{1-\bar w}-\frac{4}{3}\frac{e^2+f^2+ef}{1-\bar w^2} \\ 
h_c&=&1+\frac{2f}{1-\bar w}-\frac{4}{3}\frac{e^2+f^2+ef}{1-\bar w^2} \\ 
\bar h&=&1-\frac{4}{3}\frac{e^2+f^2+ef}{1-\bar w^2} \\ 
\Omega_{de}&=&1-\frac{4}{3}\frac{(3-\bar w)(e^2+f^2+ef)}{(1+\bar w)(1-\bar w)^2} \ . 
\eea 
These expressions agree with the numerical results for the asymptotic 
expansion rates shown in Fig.~\ref{fig:abc} to 0.03\%.  

These solutions have several interesting properties.  First, note that the 
averaged expansion rate $\bar h$ deviates from the isotropic expansion rate 
only at second order in the equation of state anisotropy.  
Second, when the $h_i$ approach 
fixed points, this means that $H_i/\hiso=$ constant, not $H_i=$ constant. 
When $\bar w\approx-1$ and so $H_{\rm iso}\approx$ constant then as long 
as the offsets $e$, $f$ are sufficiently small 
each $H_i$ is nearly constant, i.e.\ one almost 
has de Sitter-like behavior. 

This constancy of the expansion rate is reminiscent of 
the generic isotropization 
during inflation shown by \cite{wald}.  There, anisotropic matter plus a 
cosmological constant led to eventual isotropic, de Sitter expansion while 
here isotropic matter plus anisotropic dark energy leads to anisotropic 
expansion but one proportional to the isotropic case, and nearly de Sitter 
in the case that $\bar w\to-1$.  Separately, 
note that $\Omega_{de}$ goes asymptotically to a finite value 
different from 1, but 
the dimensionless matter density $\Omega_m$ still goes to 0.  The relation 
$\Omega_{de}+\Omega_m=1$ does not hold because these quantities were defined 
relative to $H_{\rm iso}$, and 
$\rho_{de}(w_a,w_b,w_c;t)\ne\rho_{de,iso}(\bar w;t)$.  

Cosmological models containing a global anisotropy, such as this Bianchi 
model, are severely constrained by observations \cite{Eriksen:2003db,Land:2005ad,Jaffe:2005pw,Hoftuft:2009rq}, specifically the 
integrated Sachs-Wolfe effect on the CMB  \cite{Campanelli:2006vb,Campanelli:2007qn,Battye:2009ze}.  
Illustratively, the temperature anisotropy arises as 
\bea 
\frac{\Delta T}{T} &\sim& \int d\eta\, \delta\dot g_{ij} \hat n^i \hat n^j 
\sim \int d\eta\,(a\dot a-b\dot b) \nonumber \\ 
&\sim& \int d\eta\,(h_a-h_b)\sim \int d\eta\, \Delta w \ , \label{eq:iswcartoon} 
\eea  
where $\eta$ is the conformal distance, $g_{ij}$ the metric, and 
$\hat n$ the line of sight 
unit vector.  More precisely, \cite{Koivisto:2008ig,Appleby:2009za} showed that 
for a dark energy model with constant equations of state 
$(\bar w+\Delta w_{a},\bar w+\Delta w_{b},\bar w - \Delta w_a- \Delta w_b)$, 
\begin{eqnarray} 
\nonumber  \frac{\Delta T}{T} &=&  -J(\Omega_{m},w) 
\left[ \Delta w_{a} \sin^{2}\theta \cos^{2}\phi + \Delta w_{b} \sin^{2} \theta \sin^{2} \phi \right. \\  & & \qquad -
\left. (\Delta w_{a} + \Delta w_{b}) \cos^{2} \theta \right] \ , 
\end{eqnarray} 
where $(\theta,\phi)$ parametrize the angular position on the sky and 
$J(\Omega_{m},w) \sim {\cal O}(1)$ is a function of the cosmological 
parameters.  This equation highlights two important points: first that 
this anisotropic dark energy model sources the CMB quadrupole only 
(to leading order in $\Delta w \ll 1$), and second that the temperature 
anisotropy is linearly proportional to $\Delta w$ (as the cartoon version 
Eq.~\ref{eq:iswcartoon} also indicated).  Therefore, barring any fine 
tuned cancellations of the leading order effect (such as through precisely 
compensated distributions of the energy-momentum, cf.\ the path integration 
over $\Delta w$ in Eq.~\ref{eq:iswcartoon}), $|\Delta w|<2\times 10^{-4}$ 
is required for this Bianchi I class of models \cite{Appleby:2009za}. 

This conclusion seems difficult to avoid.  However, let us investigate 
at what level other probes might independently constrain dark energy 
anisotropy within this model.  Also note that the CMB constraint is an 
integrated effect from recombination to the present and so using only 
the late universe might also be of interest.  
To address those issues of possible compensation (such as might arise in vector field models \cite{Koivisto:2008xf}) or time-dependent low redshift anisotropy, 
in the next section we concentrate on supernova distances, observed over 
several well separated areas of sky, such as from the deep fields of 
Dark Energy Survey \cite{des} or LSST \cite{lsst}.

%%%%%%%%%%%%%%%%%%%%%%%%%%%%%%%%%%%%%%%%%%%%%%%%%%%%%%%%%%% 
\section{Supernova Constraints on Anisotropic Expansion} \label{sec:ray} 

Type Ia supernova (SN) distances provide excellent probes of the dark energy 
equation of state in isotropic Friedmann-Robertson-Walker (FRW) universes.  
Here we apply them to an anisotropic universe such as the Bianchi I model just 
considered.  (Also see \cite{campanelli2011} for fitting current data to 
a restricted Bianchi model.)  
The supernova survey is treated as independent sky areas with 
deep, well cadenced observations suitable for accurate distance measurement. 
We consider three patches of 10 deg$^2$ each and study the effect of the 
angular distribution of the patches.  

Within each area we simulate 1000 SN with magnitudes drawn from a Gaussian 
distribution with dispersion $\sigma_m = 0.1$ and mean given by the isotropic 
expansion FRW relation with $w=-1$.  The SN are randomly distributed between 
$z=0.2-1.2$.  This gives $\sim$100 SN per 0.1 redshift bin, or a statistical 
precision of 0.01 mag per bin.  This is treated as the systematic floor, 
i.e.\ a survey may observe more SN in each patch but the effective error 
is equivalent to that of 1000 SN statistically. 

Toward each patch we solve the light propagation in the anisotropic 
cosmology using the Raychaudhuri equation.  First, the background expansion 
is given by the evolution equations 
\begin{eqnarray}  \label{eq:los2}
\dot{H}_{a} &+& H_{a}^{2} + {H_{a} H_{b} \over 2} + 
{H_{a} H_{c} \over 2} - {H_{b}H_{c} \over 2} \notag\\ 
&=& -4\pi G [P_{b}+P_{c} - P_{a}] \label{eq:17} \\ 
\dot{H}_{b} &+& H_{b}^{2} + {H_{b} H_{a} \over 2} + {H_{b} H_{c} \over 2} - {H_{a}H_{c} \over 2} \notag\\ 
&=& -4\pi G [P_{a}+P_{c} - P_{b} ] \\ \label{eq:los10}
\dot{H}_{c} &+& H_{c}^{2} + {H_{c} H_{a} \over 2} + {H_{c} H_{b} \over 2} - {H_{a}H_{b} \over 2} \notag \\ 
&=& -4\pi G [P_{a}+P_{b} - P_{c} ]  \\ 
\dot{\rho}_{m} &+& (H_{a}+H_{b}+H_{c}) \rho_{m}  = 0 \\ 
\dot{\rho}_{\rm de} &+& (1+w_{a})H_{a}\rho_{\rm de} + (1+w_{b})H_{b}\rho_{\rm de} \notag\\ 
&+& (1+w_{c})H_{c}\rho_{\rm de}  = 0 \ ,\label{eq:18} 
\end{eqnarray} 
which are basically Eqs.~(\ref{eq:littlehbegin})--(\ref{eq:littlehend}). 
These are solved starting with isotropic initial conditions 
$a = b = c = a_{\rm i}$ and $H_{a}=H_{b}=H_{c} = H_{\rm iso}$ 
at $a_{\rm i} = 2.5 \times 10^{-3}$ 
and evolved to the present, defined as $\Omega_{de,0}=0.72$.  

Once we have $H_{a,b,c}$ and $a,b,c$ 
this provides the redshift to each SN as a function of sky 
direction $z(\theta,\phi)$, and 
the Raychaudhuri equation can be used to 
determine the propagation of light rays through an arbitrary spacetime: 
\begin{eqnarray} 
\label{eq:a1} \frac{(A^{1/2})_{\lambda\lambda}}{A^{1/2}} + 
\frac{\zeta^{2}}{A^{2}} &=& -\frac{1}{2} R_{\mu\nu}k^{\mu}k^{\nu} \\ 
\label{eq:a2} \zeta_{\lambda}  &=& A \Theta \cos(\phi_\star - \phi) \\  
\label{eq:a3} \zeta \phi_{\lambda} &=& A \Theta \sin(\phi_\star - \phi) \ , 
\end{eqnarray} 
where $A^{1/2}$ is the cross sectional area of the beam, $\zeta$ the 
amplitude of the shear, and $\phi$ its phase.  Subscripts $\lambda$ 
denote derivatives with respect to the affine parameter $\lambda$, the 
photon four-momentum $k_{\mu}$ is defined by $k^{\mu} = dx^{\mu}/d\lambda$, 
$R_{\mu\nu}$ is the Ricci tensor and 
\begin{equation} 
\label{eq:p2} \Theta e^{i\phi_\star} = R_{\mu\alpha\nu\beta}k^{\mu}k^{\nu}(t^{*})^{\alpha}(t^{*})^{\beta} \ , 
\end{equation}
where $R_{\mu\alpha\nu\beta}$ is the Riemann tensor and $t^{\mu}$ is a 
complex null vector, defined via $t^{\mu}t_{\mu} = t^{\alpha}k_{\alpha}=0$ 
and $t^{\alpha}(t^{*})_{\alpha} = 1$. We use initial conditions $A^{1/2}=0$, 
$\zeta=0$.  

The area of the light ray bundle $A^{1/2}$ is linearly proportional to 
the angular diameter distance.  For an isotropic spacetime 
Eq.~(\ref{eq:a1}) reduces to the standard result 
\begin{equation} 
d_{\rm A} = \frac{\eta}{1+z}=\frac{1}{1+z} \int_0^z 
\frac{d\bar{z}}{H(\bar{z})} \ . 
\end{equation}
However, when we introduce anisotropy this relation is no longer correct 
due to the shear on the beam and the anisotropic part of the 
energy-momentum tensor (recall 
$R_{\mu\nu}k^\mu k^\nu=T_{\mu\nu}k^\mu k^\nu$). The relation 
$d_{\rm L} = (1+z)^{2} d_{\rm A}$ does hold though regardless of the 
anisotropy, and we use this to construct the luminosity distance to each SN.  
For simplicity, we use the reasonable approximation that the globally 
anisotropic (Bianchi) expansion is effectively isotropic within each 10 
deg$^{2}$ patch of the sky (i.e.\ within this $2.5\times10^{-4}$ of the 
full sky).  

Note that the redshift now contains a non-trivial angular dependence 
\bea \label{eq:los3}
1&+& z(\theta,\phi,a,b,c) = \left( \left[\frac{a(t_{0})}{a}\right]^{2} \sin^{2}\theta\cos^{2}\phi \right. \\ 
&+& \left. \left[\frac{b(t_{0})}{b} \right]^{2} \sin^{2}\theta \sin^{2}\phi  + \left[\frac{c(t_{0})}{c}\right]^{2} \cos^{2}\theta  \right)^{1/2} \ , \notag 
\eea 
so we must obtain the luminosity distances as a function of redshift in each 
patch of the sky independently. In addition, we do not set 
$a(t_{0}) = b(t_{0})=c(t_{0}) = 1$ at the present, instead we choose 
isotropic initial conditions for the scale factors. 

\begin{figure*}[htbp!]
\includegraphics[width=0.32\textwidth]{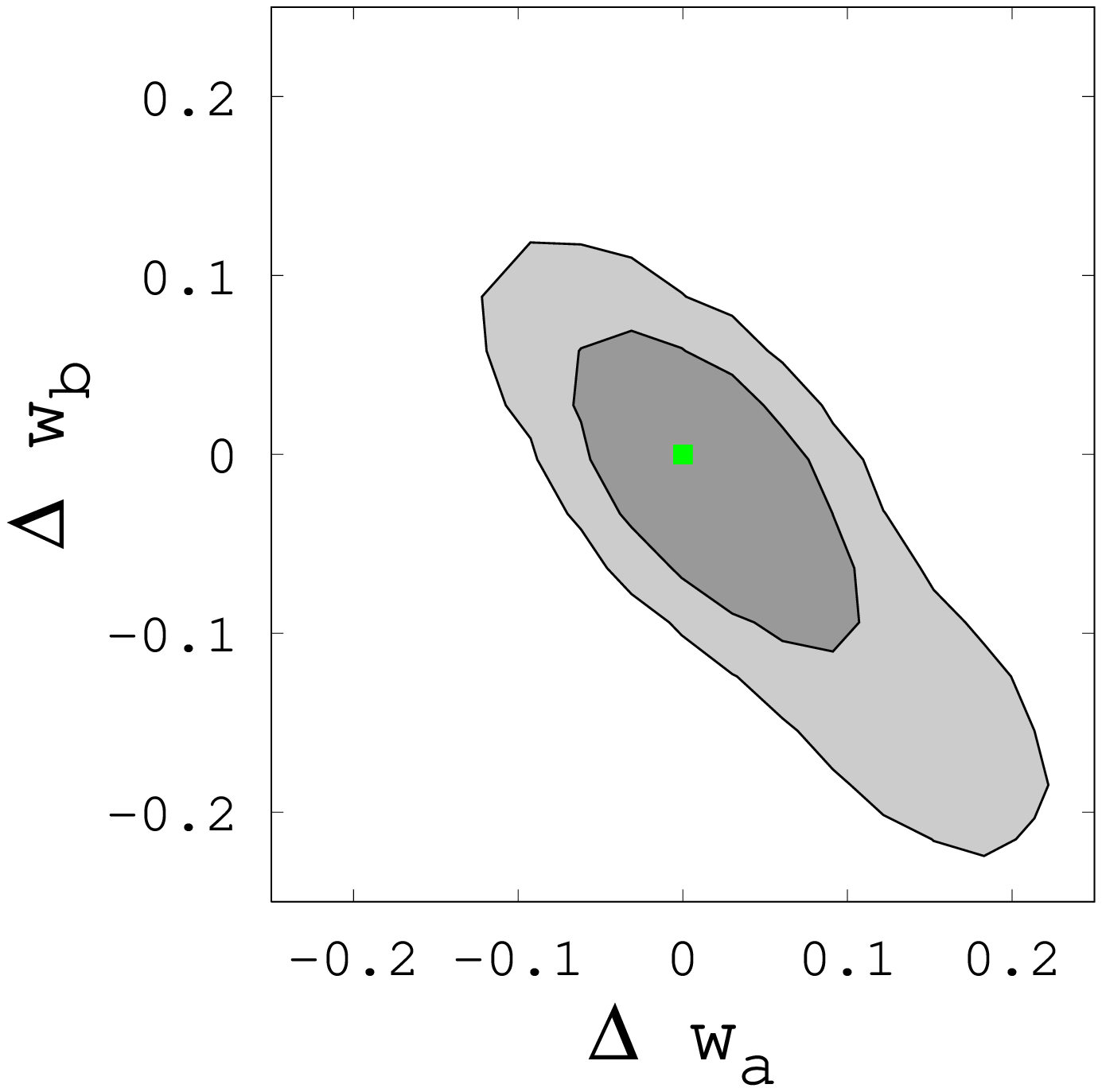}
\includegraphics[width=0.32\textwidth]{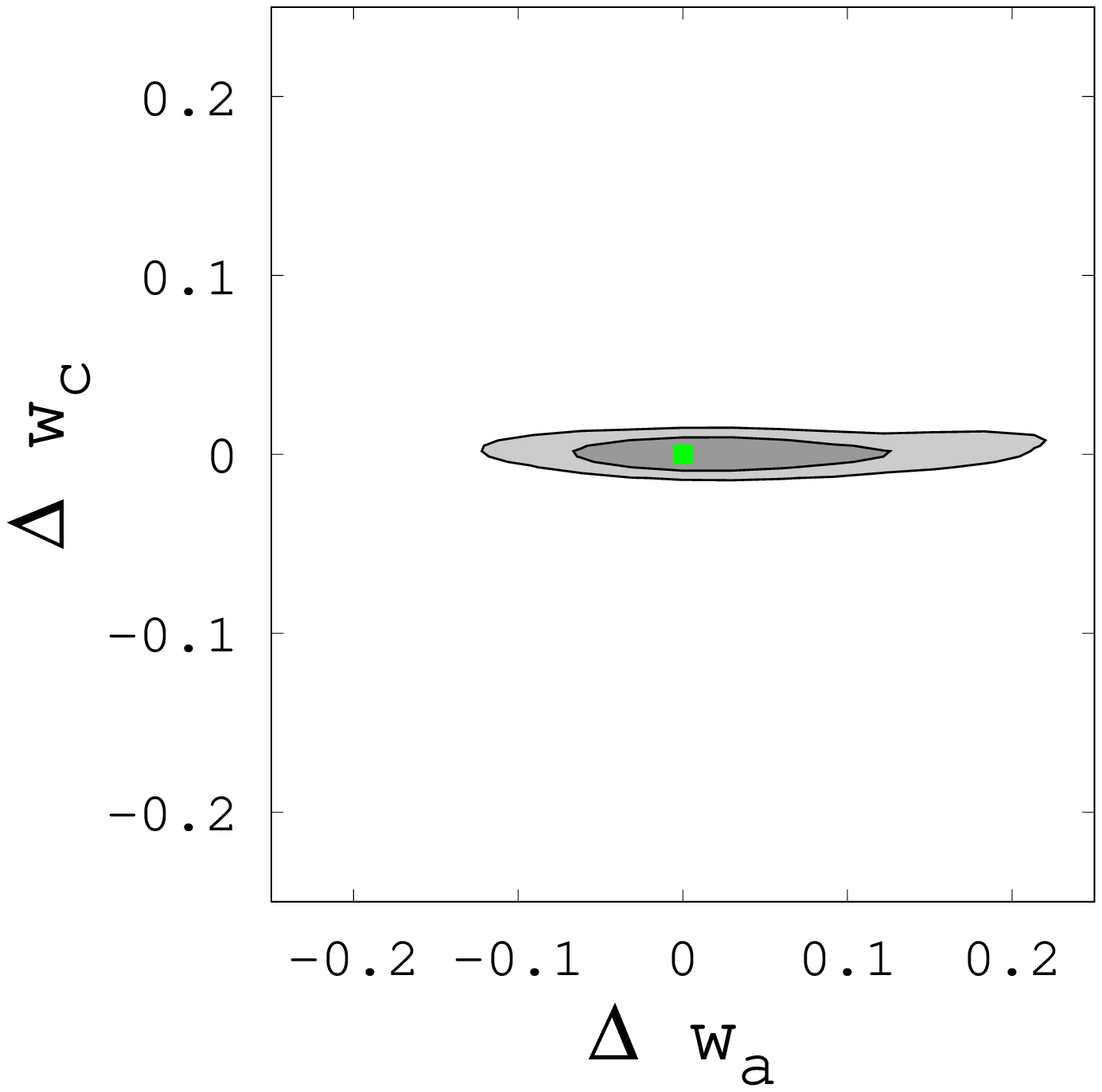}
\includegraphics[width=0.32\textwidth]{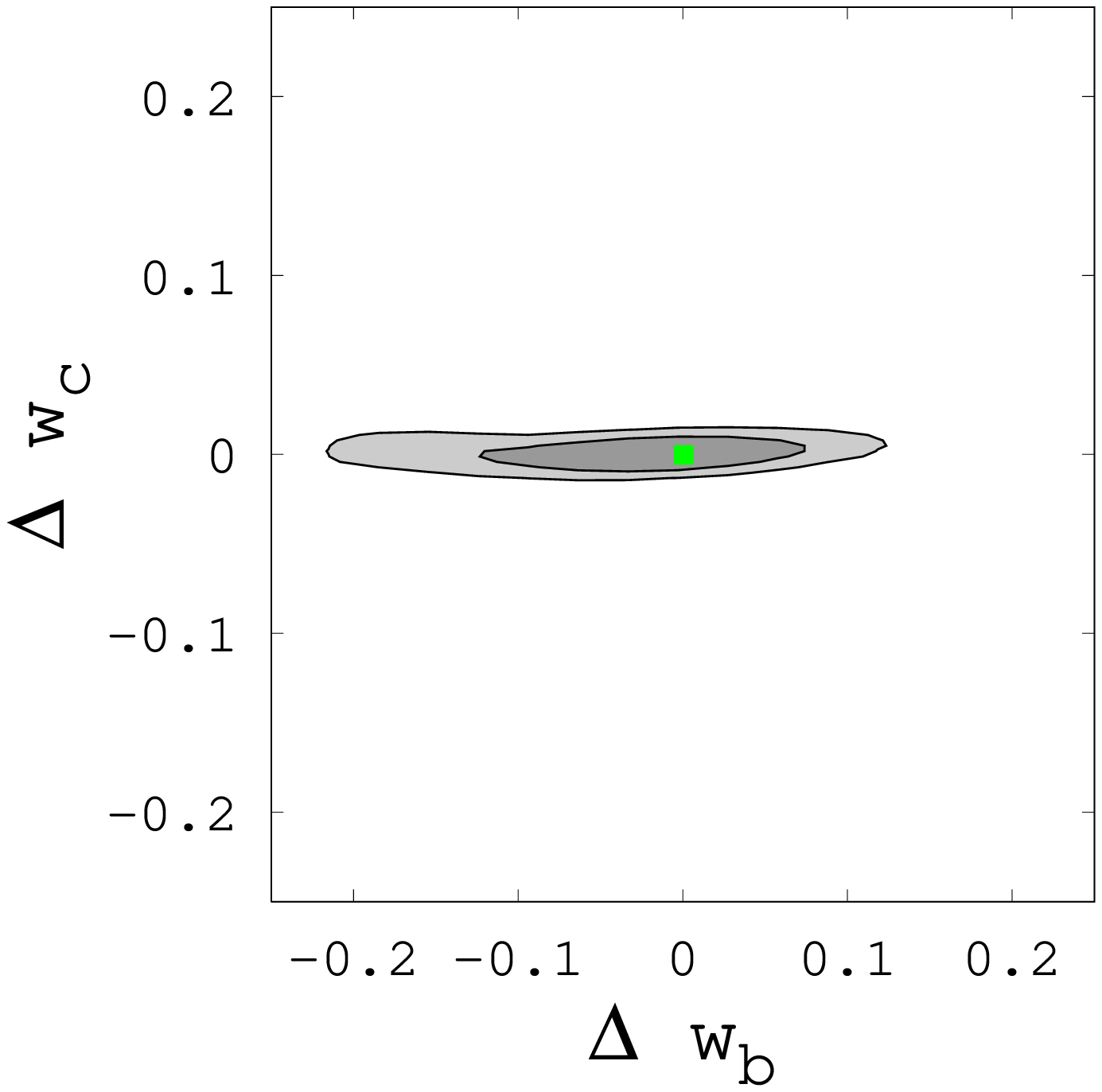} 
\caption{68\% and 95\% confidence level constraints on anisotropies 
$(\Delta w_{a},\Delta w_{b},\Delta w_{c})$ obtained through MCMC 
analysis of distance measurements are shown for the case of three patches 
in the same quadrant of the sky.  Such clustered fields yield large 
degeneracies.  The isotropic input cosmology is denoted by the green 
square. 
} 
\label{fig:n2}
\end{figure*}

\begin{figure*}[htbp!]
\includegraphics[width=0.32\textwidth]{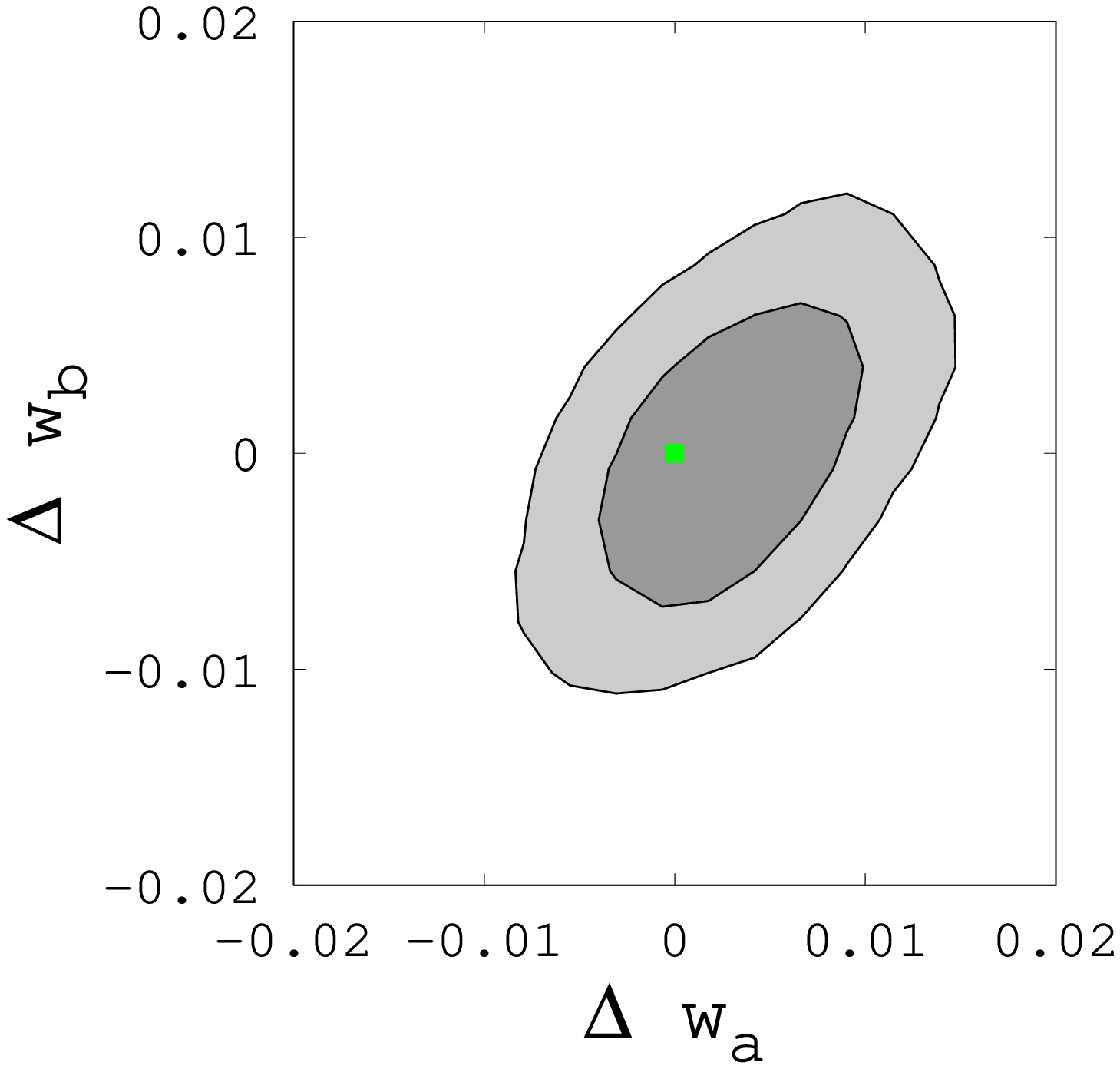}
\includegraphics[width=0.32\textwidth]{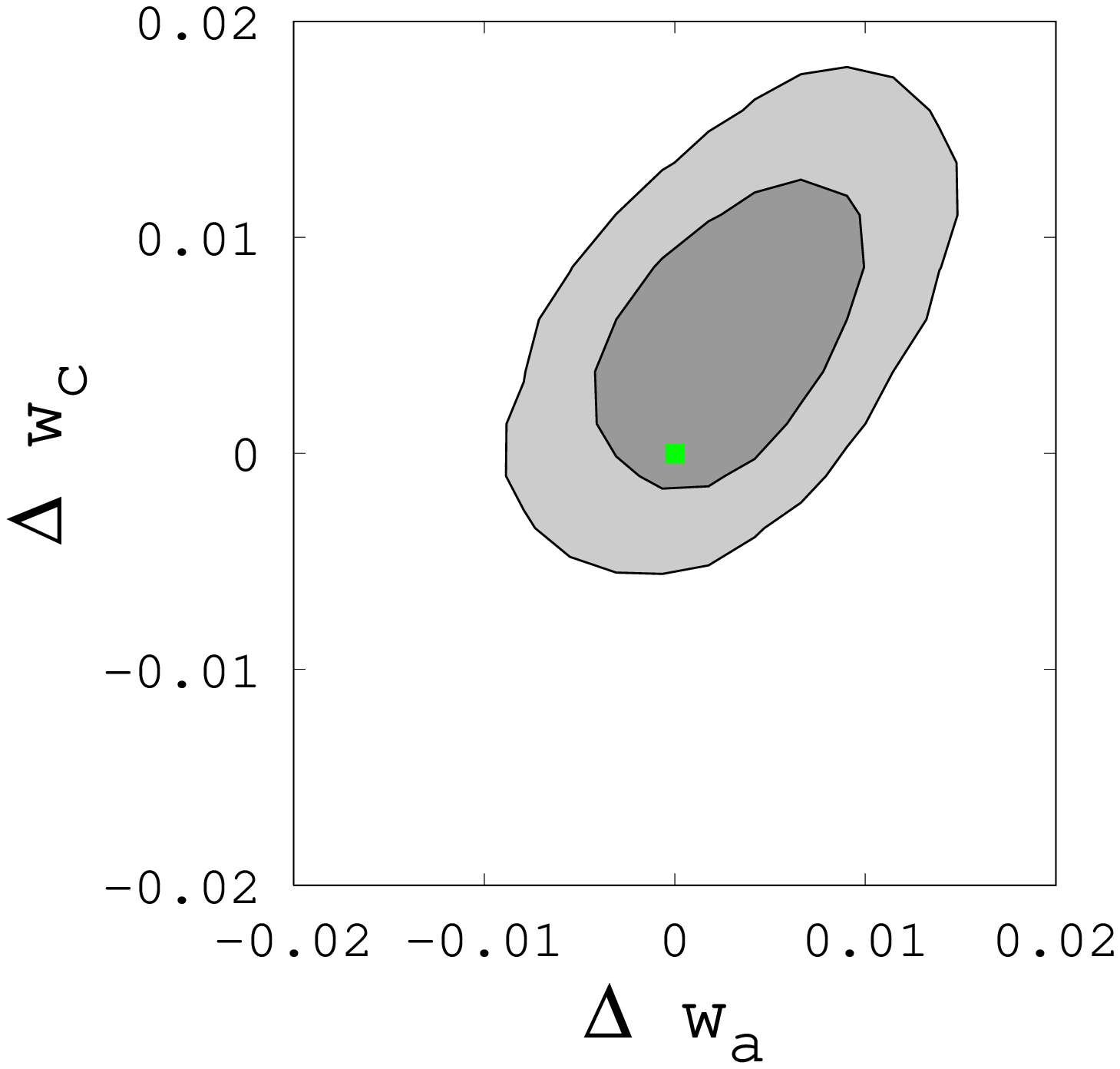}
\includegraphics[width=0.32\textwidth]{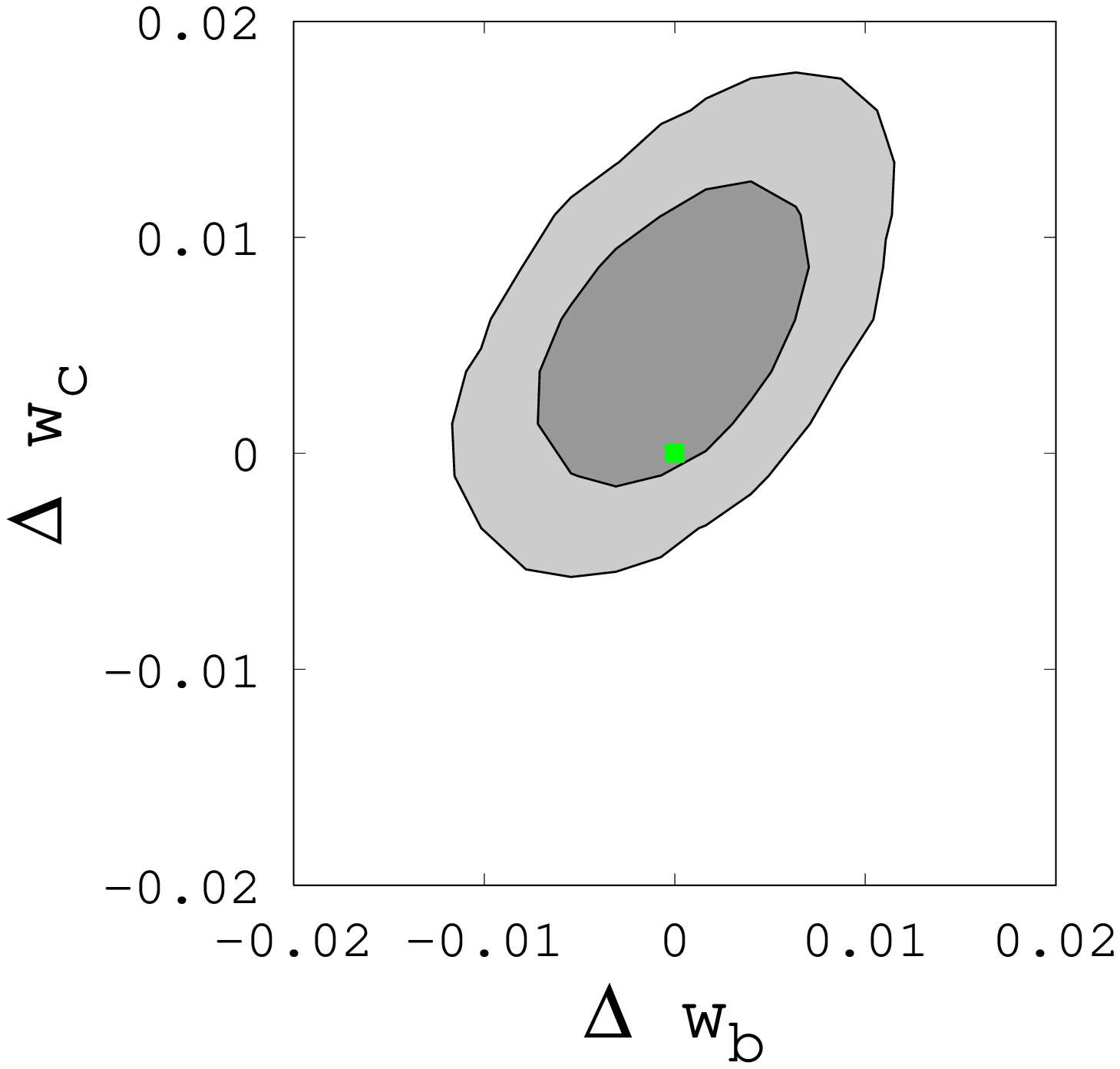}
\caption{As Fig.~\ref{fig:n2} but for the case of three patches 
in orthogonal sky directions.  Note the change in scale.  
Now the equation of state estimations are strongly constrained and 
much less degenerate. 
} 
\label{fig:n3}
\end{figure*}

We perform an MCMC analysis to confront the anisotropic model with the 
simulated SN data.  Figures~\ref{fig:n2}--\ref{fig:n3} exhibit the 
constraints placed on the dark energy equation of state anisotropy.  
We have fixed $\Omega_{de,0}=0.72$ (and always assume spatial flatness), 
both to reflect the 
constraints coming from the much wider part of the surveys (i.e.\ the 
wide fields, rather than deep SN fields) and to find the maximum constraint 
on the anisotropy. Varying over more cosmological parameters would inevitably widen the uncertainty on 
$(w_{a},w_{b},w_{c})$ and hence obfuscate our point; to find the ceiling on how well a future supernova 
experiment could constrain the anisotropy.  

In Fig.~\ref{fig:n2} we consider the patches to lie in the same quadrant 
of the sky, specifically $(\theta,\phi)=(0,0)$, $(0.15,0.15)$, 
$(0.25,0.25)$, with the angles measured in radians.  
We do not expect such a setup to be optimal for constraining 
global anisotropy; if all of the patches constrain 
$(\Delta w_{a},\Delta w_{b},\Delta w_{c})$ in a similar direction then 
degeneracies should arise.  However, surveys do sometimes select deep, 
cadenced fields within a restricted sky area.  

The optimal constraint, using fields in orthogonal directions 
$(\theta,\phi)=(0,0)$, $(0,\pi/2)$, $(\pi/2,\pi/2)$ is shown in 
Fig.~\ref{fig:n3}.  We see that the constraints are much tighter and 
less degenerate. Generically we expect maximal degeneracy between the 
equation of state parameters when the patches align in the sky, and we 
require at least three patches to ensure that the degeneracy is broken.

%%%%%%%%%%%%%%%%%%%%%%%%%%%%%%%%%%

%%%%%%%%%%%%%%%%%%%%%%%%%%%%%%%%%%%% 

We see that in the optimal case the constraints that upcoming SN surveys 
will be able to place on the global anisotropy are of order 
$\Delta w \sim {\cal O} (10^{-2})$.  This is still significantly weaker 
than the ISW bound considered in \cite{Campanelli:2006vb,Campanelli:2007qn,Battye:2009ze}. Due to the prohibitive nature 
of the CMB limit for anisotropic expansion, barring fine tuning, 
in what follows we fix the global dynamics as isotropic 
and explore possible local, line of sight effects (including those due 
to systematics).

%%%%%%%%%%%%%%%%%%%%%%%%%%%%%%%%%%%%%%%%%%%%%%%%%%%% 
\section{Line of Sight Approach to Anisotropy} \label{sec:sens} 

Going from an anisotropic theoretical model to observational predictions 
is relatively straightforward, but we often want to proceed from (possibly 
anisotropic) observations to learn about the underlying cosmology.  This 
entails some subtleties, which we begin by discussing before assessing 
the sensitivity of observational probes.  Note that one of the points of 
interest is that tests of anisotropic measurements apply not only to 
non-FRW models but to isotropic universes with anisotropic astrophysical 
systematics (such as patchy extinction and others discussed below). 

%%%%%%%%%%%%%%%%%%%%%%%%%%%%%%%%%%%%%%%%%
\subsection{Testing Isotropy and Anisotropy} \label{sec:isoaniso} 

The previous sections discussed a simple anisotropic model of dark energy, 
and considered how a future survey might place constraints on the 
cosmological parameters characterizing the anisotropy (the three 
orthogonal equations of state). Since we had a definite cosmological 
model and a closed system of equations, we were able to directly relate 
expansion observables to the cosmological parameters.

Typically however, a different approach is taken when constraining 
anisotropy.  The method in \cite{lsstbook,sullivan} for example is to observe 
different patches of the sky, and assume an FRW-like evolution in each 
direction. Specifically, the luminosity distance in each direction $\hat n$ 
is taken to be 
\begin{equation} \label{eq:los1} 
d_L(\hat n) = \frac{1+z}{H_{0}} \int_0^z \frac{dz'}{\sqrt{ \Omega_{m, 0} 
(1+z')^{3} + \Omega_{de, 0}(1+z')^{3[1+w(\hat n)]}}} \ . 
\end{equation} 
Isotropy is tested by comparing the best fit parameter values $w$ in 
each patch (usually other parameters such as $\Omega_{m,0}$ are 
taken to be direction independent). 

If the Universe (or more precisely, the data) is anisotropic, then it is 
important to realise that constraining the effective expansion history 
along a line of sight using a Friedmann equation is not a self consistent 
procedure. In the above example, if there were an anisotropic signal in 
the  expansion data (the SN distances, say) then $w$ along each line of 
sight in Eq.~(\ref{eq:los1}) does not correspond to the actual cosmological 
equation of state parameter that drives the expansion. 

One can think of this approach as a ``line of sight'' method, similar in 
spirit to the Dyer-Roeder model \cite{Dyer:1973zz} to test homogeneity.  
There, one takes a globally Friedmann expansion history but posits that along 
certain lines of sight the light bundles will feel a different matter 
distribution.  In Eq.~(\ref{eq:los1}) one also assumes a globally Friedmann 
expansion, and yet allows $w$ to vary with direction. This is an acceptable 
procedure as a consistency test of whether the isotropic FRW cosmology can fit 
the data.  However to explore anisotropic models, and robustly deal with 
anisotropic signals in the data, one must find a way of relating the 
purely phenomenological $w(\hat n)$ in Eq.~(\ref{eq:los1}) to the physical 
expansion (i.e.\ the actual equation of state) in the proposed anisotropic 
model. 

For the Bianchi I spacetime, the connection between the anisotropic 
distance-redshift relation and the dark energy equation of state is 
straightforward; it is provided by Eqs.~(\ref{eq:los2}-\ref{eq:p2}).  
Note that even if we can relate $w(\hat n)$ to an actual cosmology, we 
still cannot generically use the standard relationship Eq.~(\ref{eq:los1}). 
This expression does not take into account the redshift angular dependence 
$z=z(\theta,\phi)$ of Eq.~(\ref{eq:los3}) or the beam shear that alters the 
angular diameter distance in the presence of an anisotropic fluid component. 
For astrophysical origins of anisotropy (see the next subsection for 
examples), adjustments must often be made quite early in the data analysis, 
e.g.\ extinction corrections enter in the lightcurve parameter fitting stage 
for SN rather than in the final distances. 

Given the above issues, two questions should be addressed concerning the 
line of sight approach: 1) False positives -- if the data is genuinely 
isotropic how accurately will the analysis be able to verify this and 
constrain 
anisotropies?, 2) False negatives -- if the data is 
actually anisotropic, how accurately will the analysis be able to measure 
this, and rule out isotropy, 
given that the method is only consistent for isotropic data?  
Finally, if the method behaves well enough that we accept its formal 
shortcomings, then how sensitive are the various late time cosmological 
probes to anisotropies in the data. 

The first question can be addressed by populating our mock supernova 
sample using an isotropic cosmological model, and then performing an MCMC 
analysis using the full Bianchi machinery to fit $(w_a,w_b,w_c)$ 
of the spacetime, or using 
the line of sight approach to fit $(w_1,w_2,w_3)$ 
of the patches, and testing each for 
isotropy.  The relative magnitudes of the errors obtained using the two 
methods will inform us as to the reliability of the line of sight approach.  
The input cosmology is $\Lambda$CDM and we use similar SN data 
characteristics as in Sec.~\ref{sec:ray}.  
Both approaches reproduce the input cosmology, as expected, and the 
errors are of the same order of magnitude, although the line of sight 
approach gives $\sim50\%$ larger uncertainties on $\Delta w$ 
($6\times 10^{-3}$ rather than $4\times 10^{-3}$, likely due to treating 
the parameters as independent in each field).  
We conclude that the line of sight approach is a viable 
method of testing isotropic data, despite the fact that it does not 
consistently take into account cosmological anisotropy. 

\begin{figure}[htbp!]
\includegraphics[width=\columnwidth]{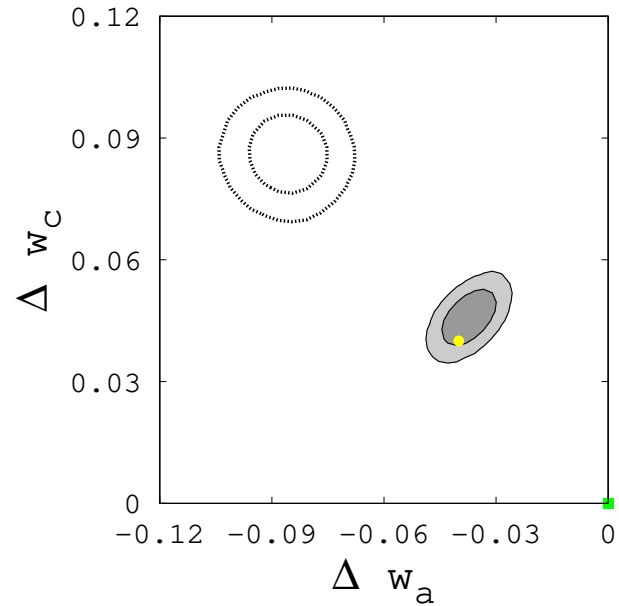} 
\caption{68\% and 95\% CL contours are presented for fitting for an 
anisotropic input cosmology when solving the full Raychaudhuri cosmological 
equations (gray shaded contours) and when using the line of sight approach 
(unfilled dotted contours).  Both approaches accurately reject 
the isotropic, $\bar w=-1$ case (green square) and the Raychaudhuri 
method recovers the input cosmology (yellow dot).  The line of sight method 
actually constrains combinations of the $w_i$ (but this is not realized 
without knowing the true cosmology).  The other 2D projections not shown 
look similar.  
}
\label{fig:2D} 
\end{figure}

The second question, that of false negatives, i.e.\ deriving isotropy 
spuriously because of using an (isotropic) FRW expression for distance, 
can be addressed by populating three patches in the sky using an anisotropic cosmological model, and then performing an MCMC analysis of the parameter space for the two different approaches. Specifically, we use the full Bianchi I equations to construct the magnitudes of 3000 supernovae in three orthogonal 
patches in the sky, using equation of state parameters 
$(\Delta w_{a},\Delta w_{b},\Delta w_{c}) = (-0.04,0,0.04)$, with respect to 
$\bar w=-1$.  We then employ the full Bianchi I equations in the first approach, and the line of sight equations in the second. 
Figure~\ref{fig:2D} exhibits the results. The gray shaded confidence 
contours are 
obtained using the full anisotropic equations; as expected the best fit is very close to the input cosmology and we are able to distinguish this model from isotropic $\Lambda$CDM at high confidence. 

The contours corresponding to the line of sight approach are presented as 
dashed lines; here we see a significant bias in the best fit value obtained 
in the analysis. This is due to the fact that the Hubble parameters 
$H_{\rm a,b,c}$ along each line of sight are not simply sourced by 
$w_{\rm a,b,c}$ individually and independently, but rather by linear 
combinations of them (see Eqs.~($\ref{eq:los2}-\ref{eq:los10}$)).  
Hence we are effectively constraining $(w_{\rm b} + w_{\rm c} - w_{\rm a},w_{\rm a} + w_{\rm c} - w_{\rm b},w_{\rm a} + w_{\rm b} - w_{\rm c})$, 
though we only realize that by using the Bianchi analysis not the Friedmann, 
and hence the best fit is biased.  
This is not the only difference between the methods however; the line of sight approach also does not take into account anisotropic effects such as beam shear or the 
non-trivial relationship between $z$ and $a,b,c,\theta,\phi$. These differences account for the fact that the errors obtained using the two methods are different, and the line of sight approach yields perfectly non-degenerate contours. 

In spite of the problems with the line of sight approach, it is clear that 
if there is anisotropy in the data then the method should detect it.   
That is, no triplet of the linear combinations of $w_i$ will have all the 
same elements unless all individual $w_i$ are identical, so false negatives 
are avoided.  
How one interprets the anisotropy signal without knowing the underlying 
cosmological model is not clear however. In this work, we can roughly 
relate the line of sight method to cosmological parameters since we have 
created the data using a specific anisotropic model.  With real data, we 
no longer have the luxury of knowing the source of the anisotropy.

There is one final effect that must be considered. In the above analysis 
we have taken the supernova deep fields to lie in orthogonal directions. This will provide a maximal constraint on the anisotropy of the data, however it is also expected to be the setup for which the two approaches will have 
closest agreement. This is due to the fact that in the line of sight approach, we are assuming that the directional dependent equation of state parameters are uncorrelated. However, if the fields are all located in the same region, then we expect an additional deviation between the two methods as a result 
of the correlation between the fields' equations of state, although such 
fields will also deliver poorer constraints.

%%%%%%%%%%%%%%%%%%%%%%%%%%%%%%%%%%%%%%%% 
\subsection{Sensitivity to Anisotropy} \label{sec:lossens} 

The line of sight approach is therefore adequate for testing isotropy 
and (the presence of) anisotropy.  Moreover, it 
permits exploration not only of anisotropy from the cosmological model 
but from astrophysical systematics.  
For example, measurements of supernova 
distances in directions with different extinctions would imply different 
cosmological parameters for the distance-redshift relation if the patchy 
extinction was not fully recognized.  Indeed, at the levels of accuracy 
required for future distance measurements, work is still ongoing in 
mapping inhomogeneous dust extinction in our Milky Way galaxy 
\cite{10124804}.  Another example is baryon acoustic oscillation (BAO) scale 
distances measured through galaxy clustering.  Anisotropic stellar 
density can either obscure or augment the galaxy clustering correlation 
function if not fully recognized \cite{12036499}; indeed before correction 
this gives a $2.6\sigma$ difference between the BAO scale measured from 
Northern Galactic Cap and Southern Galactic Cap fields (see Appendix A of 
\cite{12036594}).  Other possible astrophysical anisotropies include 
a locally anisotropic electron optical depth in CMB measurements (e.g.\ see 
\cite{10020836} and references therein) and patchy reionization, which 
can affect CMB, 21 cm, and even BAO cosmology inferences 
\cite{0702099,0511141,9805012,0503166,0604358}.  

The question we consider now is how sensitive various late time 
cosmological probes are to any such anisotropy, and over which redshifts.  
We emphasize that $\Delta w$ is merely a proxy, a common language, for 
comparing such sensitivities, and may have nothing to do with a physical 
equation of state.  The probes considered are 
the distance-redshift relation $d(z)$, 
e.g.\ as measured through Type Ia supernovae or baryon acoustic oscillations, 
the Hubble parameter $H$, e.g.\ through radial BAO, and the reduced 
distance to CMB last scattering $d_{lss}$.  We also consider probes of 
growth variables such as the growth factor $g=D/a=(\delta\rho_m/\rho_m)/a$ 
normalized to one at high redshift, e.g.\ as measured from weak gravitational 
lensing or galaxy surveys, and the growth rate $f=d\ln D/d\ln a$ in the 
products $f\sigma_8(z)\sim fD$ and $f\sigma_8/\sigma_{8,0}$, calibrated 
to high redshift and low redshift, respectively, e.g.\ from redshift space 
distortions. 

\begin{figure}[htbp!]
\includegraphics[width=\columnwidth]{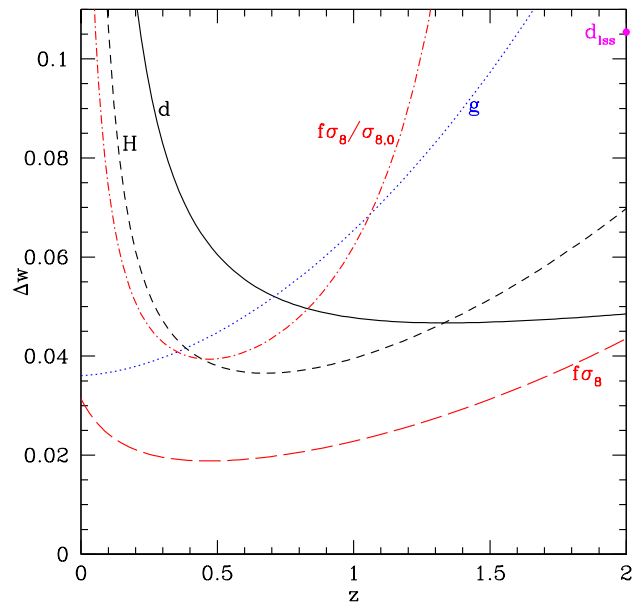}
\caption{Each curve represents the sensitivity $\Delta w$ to dark 
energy anisotropy made possible by 1\% measurements of the labeled 
observable, as function of measurement redshift.  The CMB $d_{lss}$ 
sensitivity is shown on the right axis by the purple filled circle. 
}
\label{fig:sens}
\end{figure}

Figure~\ref{fig:sens} exhibits the sensitivities to anisotropies 
$\Delta w$ between lines of sight as a function of the redshift $z$ of 
the measurement, for 1\% accuracy on different observable quantities 
${\mathcal O}$.  That is, 
\be 
\Delta w_{1\%}=\left(\frac{\partial{\mathcal O}}{\partial w}\frac{1}{0.01{\mathcal O}}\right)^{-1} \ . 
\ee  
Again, $\Delta w$ means that level of variation in the observable from 
any anisotropy source equivalent to a change $\Delta w$.  
Seeing anisotropies that have smaller $\Delta w$ than in the Figure 
would require better than the 1\% measurement 
accuracy.  The assumption here is that this is a differential measurement 
on the sky, 
and the overall wide field survey determines the background values of 
all other cosmological parameters.  Thus the figure 
gives lower limits on the sensitivity to anisotropy $\Delta w$ between 
different lines of sight.  

One must fold into the figure the level of accuracy which a 
particular observable quantity would actually attain. 
For example, the CMB distance 
$d_{lss}$ may be measured by the Planck satellite to 0.2\% \cite{planck}, 
while $H$ is generally measured less well than $d$ from BAO.  Furthermore, 
the precisions must be scaled to reflect the area of sky used to compare 
lines of sight.  The 0.2\% precision for $d_{lss}$ is for full sky, but 
to look for anisotropy one must split up the area into patches, so the 
precision would degrade.

For some probes the angular scales of sensitivity to anisotropy are 
limited by the nature of the observable.  For example, both CMB acoustic 
peaks and BAO have angular sizes of $\sim1$ degree, so they lose sensitivity 
to anisotropies on smaller scales.  On the other hand, supernovae or weak 
lensing, for example, can probe down to smaller scales.  
We expect higher derivative quantities 
such as growth rates relative to growth, or the Hubble parameter relative to 
distance, to be less accurately measured.  

Taking these various factors into account, from Fig.~\ref{fig:sens} we 
anticipate that the most sensitive probe of such anisotropy will be supernova 
distance measurements, with possibly low redshift growth factor 
measurements from weak lensing and the growth rate from redshift space 
distortions contributing, especially to small scale anisotropy constraints. 
Large surveys, both spectroscopic and photometric, play roles in 
constraining dark energy anisotropy (including through determining the 
other background quantities).  Photometric errors 
propagate through to roughly the same errors in distance, i.e.\ 
\be 
\frac{\Delta d_l}{d_l}=\frac{\Delta z}{1+z}\,\left[1+\frac{(1+z)^2}{Hd_l}\right]\approx \frac{\Delta z}{1+z}\ , 
\ee 
so as long as photometric errors in a redshift bin composed of many 
objects can be constrained well, the distance uncertainties will be 
controlled sufficiently to allow testing anisotropy.  Thus, a wide field 
galaxy, or supernova, survey such as LSST could be used to investigate 
anisotropic properties of dark energy, as studied empirically in 
\cite{lsstbook}.

%%%%%%%%%%%%%%%%%%%%%%%%%%%%%%%%%%%%%%%%%%%%%%%%%%%%%%%%% 
\section{Conclusions} \label{sec:concl}

The cosmic microwave background radiation delivers strong evidence for 
isotropy, restricting global anisotropy to the $\sim10^{-5}$ level.  
This severely disfavors anisotropic models such as a Bianchi I universe. 
Lower redshift wide field surveys can deliver constraints at the percent 
level.  
Preserving isotropic expansion dynamics but allowing for local anisotropy 
remains a possibility, at least on a phenomenological level.  This Ansatz 
is similar to that of the Dyer-Roeder model, where global dynamics can 
stay Friedmann-Robertson-Walker despite lines of sight having differing 
properties.  

We have calculated exact solutions of the anisotropic Bianchi I cosmology 
and shown that even in the case of extreme anisotropy the expansion can 
retain FRW-like characteristics.  Indeed, the expansion rate in different 
directions does not have to diverge, but can go to fixed points.  We give 
analytic expressions for these through second order in the dark energy 
equation of state anisotropy.  The average expansion rate equals the 
expansion rate of the associated FRW universe at first order.  

Carrying out Monte Carlo simulations of deep fields within a wide field 
survey, {\`a} la Dark Energy Survey or LSST, we study the effect of 
the configuration of deep field distance measurements on the 
global anisotropy constraints.  Sky areas that are well separated in 
orthogonal directions break degeneracies and give tight constraints. 

Adopting a phenomenological Ansatz with direction dependent pressure (or 
equation of state) but 
global isotropy requires careful thought.  However, the results of our 
Bianchi I analysis help motivate that an Ansatz retaining a 
globally isotropic expansion could serve as a reasonable approximation, 
and our Monte Carlo results show that the line of sight approach, handled 
carefully, can give consistent results for isotropy or an alarm for 
anisotropy (including astrophysical systematics). We stress that when 
using the line of sight approach, one cannot interpret an anisotropic 
signal in terms of cosmological parameters in a straightforward manner. 

We then investigated the constraints that different astrophysical observations 
could place on such anisotropy.  For small angular scales, supernova 
distances and redshift space distortions have good leverage, while on 
large angular scales BAO and CMB distances impose limits.  Both spectroscopic 
and photometric surveys can contribute constraints, with next generation 
surveys capable of limiting anisotropies (described in the proxy language 
of dark energy equation of state $\Delta w$) 
at the $\sim5\%$ level at each redshift (with tighter constraints from summing 
over a redshift range). 

We emphasize several caveats.  A definite model for anisotropic 
dark energy that preserves isotropic expansion to the level required by 
the CMB requires further work.  Standard inhomogeneous perturbations, 
from a low sound speed for example, do not suffice.  The pressure 
perturbations may be decoupled though from the density ones by adopting an 
infinite sound speed such as in the cuscuton model \cite{cuscuton}.  
Large surveys give 
strong constraints but must be subdivided into patches to compare the 
equation of state along different lines of sight, diluting their 
effective volume.  We have outlined a number of systematics that are 
direction dependent, such as 
patchy extinction or gravitational lensing, and could give spurious signals 
for line of sight variation.  This article demonstrates some interesting 
features and results regarding testing dark energy anisotropy but 
also applies, probably more realistically, to astrophysical systematics. 

\acknowledgments

We thank Richard Battye, David Rubin, David Schlegel, Tristan Smith, and 
Hu Zhan for useful discussions.  
This work has been supported by World Class 
University grant R32-2009-000-10130-0 through the National Research 
Foundation, Ministry of Education, Science and Technology of Korea, and 
in part by the Director, 
Office of Science, Office of High Energy Physics, of the U.S.\ Department 
of Energy under Contract No.\ DE-AC02-05CH11231.

%%%%%%%%%%%%%%%%%%%%%%%%%%%%%%%%%%%%%%%%%%%%%%%%%%%%%%%%%%%%

\end{document}